\documentclass[manuscript,screen]{acmart}
\AtBeginDocument{%
  \providecommand\BibTeX{{%
    \normalfont B\kern-0.5em{\scshape i\kern-0.25em b}\kern-0.8em\TeX}}}

\setcopyright{acmlicensed}
\copyrightyear{2024}

\acmConference[CHI 2024]{LLM-BASED SYNTHETIC PERSONAE AND DATA IN HCI - Workshop}{Honolulu, Hawai'i}
\begin{document}
\title{Concerns on Bias in Large Language Models when Creating Synthetic Personae}

\author{Helena A. Haxvig}
\email{helenaamalie.haxvig@unitn.it}
\orcid{0009-0003-3858-6617}
\affiliation{%
  \institution{Dipartimento Di Ingegneria E Scienza Dell'Informazione, Università Di Trento}
  \city{Povo}
  \state{Autonomous Province of Trento, TN}
  \country{Italia}
  \postcode{38123}
}


\begin{abstract} \parskip=-0.01pt
    This position paper explores the benefits, drawbacks, and ethical considerations of incorporating synthetic personae in HCI research, particularly focusing on the customization challenges beyond the limitations of current Large Language Models (LLMs). These perspectives are derived from the initial results of a sub-study employing vignettes to showcase the existence of bias within black-box LLMs and explore methods for manipulating them. The study aims to establish a foundation for understanding the challenges associated with these models, emphasizing the necessity of thorough testing before utilizing them to create synthetic personae for HCI research.
\end{abstract}


 

\received{22/02/2024}
\received[accepted]{05/03/2024}

\maketitle 

\section{Introduction} 
Incorporating Large Language Models (LLMs) as synthetic personae in the evolving landscape of Human-Computer Interaction (HCI) research presents both interesting opportunities, daunting challenges and concerns that warrant careful consideration about critical concerns of bias and other flaws in LLMs \cite{rozado_political_2023, haxvig_exploring_2023,wan_kelly_2023}. One immense concern relates to the existence of bias in the models, and creating synthetic personae has the potential to aid the investigation of how different forms of bias manifest in LLMs, by introducing a new method of testing. However, the black-box nature of a majority of these models, and their inability to express 'opinions' contrary to overall LLM rules or fail-safes,  introduces complexities in how to prompt the models to act out specific synthetic personae in various scenarios. 

This position paper introduces an exploration of a few fundamental questions: What are the benefits and drawbacks of using synthetic personae in HCI research, and how can we customize them beyond the limitations of current LLMs? The perspectives presented in this paper have sprung from the sub-study of a PhD project on Artificial Intelligence and Participatory Design \cite{simonsen_routledge_2013}. The sub-study, currently a work in progress, aims at developing a novel method of adversarial testing \cite{dairai_adversarial_2023, mokander_auditing_2023, xu_llm_2023} through the use of contextualized "real-life" vignettes \cite{barter_use_1999, sampson_turning_2020} prompted to the interfaces of multiple LLMs to identify potential bias, trying to open up the "black box" from a more qualitative human-computer interaction perspective \cite{haxvig_exploring_2023}.

\section{Bias Detection in LLM Interfaces}
Research in various sub-fields has shown that human engagement in AI design, development, and evaluation, particularly in a qualitative manner, can ensure a focus on the socio-technical embeddedness of AI \cite{cherrington_features_2020}. This can help include socio-behavioral attributes to improve contextual understanding and interoperability, or identify potential traps developers might fall into by proactively detecting issues and ethical risks during the development process \cite{papakyriakopoulos_qualitative_2021}. 

In alignment with this, the present sub-study focuses on conducting a pilot study employing vignettes as a new method to showcase the existence of bias within black-box Language Model Models (LLMs) and exploring methods to stress the models through enactment of personae. Emphasizing the necessity of thorough testing before utilizing these LLMs to create synthetic personae, the study aims to establish a foundation for understanding the challenges associated with these models. Furthermore, the research is particularly attentive to Feminist and Queer HCI \cite{bardzell_feminist_2010, scheuerman_how_2019, devito_queer_2021, strengers_adhering_2020} considerations, acknowledging the importance of a critical stance in understanding and possibly mitigating biases in LLMs for the responsible creation of synthetic personae.

The sub-study began with pilot tests to determine which LLM interfaces are most suited for the study, culminating in the development of a systematic strategy for the vignette tests. The pilot tests explored various approaches to prompt engineering and adversarial testing methods to explore the malleability, susceptibility to specific prompts, and limitations of LLMs.

\subsection{Pilot Testing with Adversarial Attacks}
The pilot study initially aimed to assess some of the largest and most prominent LLMs existing today, considering factors such as availability, commercialized online interfaces, and prototype accessibility. The study included interfaces such as ChatGPT 3.5 turbo, Google BARD (using PaLM 2 until Gemini 1.0's launch in February 2024), Gemini, PI.ai (Inflection-1), and Coral (Cohere model). Additionally, prototype testing was conducted on Falcon 180B, LlaMa 2 70B, Guanaco 33B, and Vicuna 33B.

Existing research on bias in AI training data \cite{crawford_atlas_2021, eubanks_automating_2019}  and recent investigations into bias in Large Language Models (LLMs) highlight the potential risks of bias manifestation in LLMs \cite{rozado_political_2023, wan_kelly_2023}. The initial phase, thus, involved 'interviewing' the models on bias in LLMs and awareness of potential flaws like hallucinations. When directly questioned about bias, most models acknowledge the possibility, citing concerns related to gender, ethnicity, culture, religion, politics, ability, and age. While many models assert their attempts to maintain impartiality, some, like ChatGPT 3.5, Gemini, and Cohere, elaborate on the origins of bias, attributing it to training data, sampling bias, algorithmic bias, confirmation bias, and leading questions.. This initial testing, comprised of leading questions to assess the general embedded rules on inappropriate behavior, revealed no significant differences between the models. Further testing, involving adversarial attacks inspired by examples from DAIR.AI \cite{dairai_adversarial_2023}, assessed logical reasoning, resistance to prompt injection, and resistance to jailbreaking techniques, including creative prompts like playing a game or enacting the DAN (Do Anything Now) character for illegal activities among others. This provided some noteworthy insights, particularly in exploring the models' abilities to assume different personae. Some models resisted DAN manipulation for illegal instructions but exhibited potential for expressing biases, such as racial and gender bias, when instructed to embody specific personae. Not all models succumbed, but those that did show promise in adopting positive characters. Only two models, PI and Vicuna, were willing to adopt offensive behavior with a basic jailbreaking prompt. 

This presents a challenge in creating synthetic personae as the models respond differently to the same prompts, even if they share a similar cautious "personality". As such, it is necessary to determine whether a relatively universal approach to synthetic personae is feasible or if unique prompts are required for each model. Additionally, addressing models resistant to manipulation poses a challenge in creating heterogeneous synthetic personae. And, when stressing the models with different approaches we further risk creating situations where the model is escaping control, which would be critical in e.g. a workshop with human participants.

Some of these challenges will be explored and addressed in the subsequent steps of the sub-study, where the idea is to combine the vignette technique with ideas from adversarial attacks. Scenarios and personae will be built on the basis of empirical interview data and existing literature, and these will be prompted to the LLMs' interfaces. This allows the LLMs to operate based on these personae's perspectives and respond to presented scenarios. While these personae are crafted through research, instructing the models to embody them could result in a synthetic persona shaped by the models' inherent biases. This can produce valuable insights into how bias manifests in these models and explore strategies for how we can move beyond the limitations of LLMs when prompting synthetic personae.

\section{Ontological and Ethical Concerns}

Technological development does not happen in a vacuum and technologies are not simply passive tools, but social interventions that require engagement in moral discourse \cite{frauenberger_ways_2019}. With the inclusion of a few points that warrant further discussion, this section underscores the need for a thoughtful and ethical approach to incorporating LLMs in various contexts, emphasizing the importance of responsible design practices.

In a time where the words we apply to identify ourselves have become more open to interpretation, language serves as an imperfect reflection of shifting social realities  \cite{kaltheuner_fake_2021}, which begs us to question whether reducing the human experience to classifications in LMMs produces adequate imitations of said realities. The lack of a deep understanding of real-world contexts, cultural nuances, and human emotions in LLMs raises concerns about their ability to accurately represent personae, not to mention diverse user experiences, in Human-Computer Interaction (HCI). This is a particular concern when creating synthetic personae from potentially flawed and biased "black box" systems. In areas like Participatory Design \cite{simonsen_routledge_2013}, where amplifying marginalized voices is paramount, synthetic personae must be instruments for empowerment rather than biased obstacles. 

Lastly, conducting experiments with LLM-generated synthetic personae, especially in dynamic real-world scenarios involving humans, poses risks and requires rigorous vetting for potential harm and unpredictability before deployment. As we navigate the landscape of LLMs and HCI, it is imperative to approach the topic with ethical responsibility and critical scrutiny, exploring how to test a model's suitability before using it to create synthetic personae.

\section{Future Work}
At the current point in time, the pilot tests have been carried out and provided insights relevant for the strategy of the next steps. Now, the focus will move to creating the mentioned vignettes and "interviewing" the LLMs to test their articulation of bias, particularly on feminist and queer rights issues.
In addition to developing this innovative interview method for exploring LLMs' portrayals of sensitive topics (i.e. inherent bias), this study also aims to establish a workshop method with LLMs as non-human participants (i.e. synthetic personae) as a novel non-anthropocentric approach for semi-structured adversarial testing of bias articulation in LLM interfaces, in alignment with principles of more-than-human design approaches \cite{coulton_more-than_2019, loi_co-designing_2019}.
The current sub-study is expected to be followed with a speculative design approach, envisioning training LLMs on specifically selected data, e.g. with contrasting worldviews to provoke critical discussions about embedded values in technology. This provotyping could challenge prevailing representations and prompt us to consider how creating specific synthetic personae can guide HCI research into LLM behaviour and human-LLM interaction.

\bibliographystyle{ACM-Reference-Format}
\bibliography{LLM.workshop}


\begin{thebibliography}{21}


\ifx \showCODEN    \undefined \def \showCODEN     #1{\unskip}     \fi
\ifx \showDOI      \undefined \def \showDOI       #1{#1}\fi
\ifx \showISBNx    \undefined \def \showISBNx     #1{\unskip}     \fi
\ifx \showISBNxiii \undefined \def \showISBNxiii  #1{\unskip}     \fi
\ifx \showISSN     \undefined \def \showISSN      #1{\unskip}     \fi
\ifx \showLCCN     \undefined \def \showLCCN      #1{\unskip}     \fi
\ifx \shownote     \undefined \def \shownote      #1{#1}          \fi
\ifx \showarticletitle \undefined \def \showarticletitle #1{#1}   \fi
\ifx \showURL      \undefined \def \showURL       {\relax}        \fi
\providecommand\bibfield[2]{#2}
\providecommand\bibinfo[2]{#2}
\providecommand\natexlab[1]{#1}
\providecommand\showeprint[2][]{arXiv:#2}

\bibitem[Bardzell(2010)]%
        {bardzell_feminist_2010}
\bibfield{author}{\bibinfo{person}{Shaowen Bardzell}.} \bibinfo{year}{2010}\natexlab{}.
\newblock \showarticletitle{Feminist {HCI}: taking stock and outlining an agenda for design}. In \bibinfo{booktitle}{\emph{Proceedings of the {SIGCHI} {Conference} on {Human} {Factors} in {Computing} {Systems}}} \emph{(\bibinfo{series}{{CHI} '10})}. \bibinfo{publisher}{Association for Computing Machinery}, \bibinfo{address}{New York, NY, USA}, \bibinfo{pages}{1301--1310}.
\newblock
\showISBNx{978-1-60558-929-9}
\urldef\tempurl%
\url{https://doi.org/10.1145/1753326.1753521}
\showDOI{\tempurl}


\bibitem[Barter and Renold(1999)]%
        {barter_use_1999}
\bibfield{author}{\bibinfo{person}{Christine Barter} {and} \bibinfo{person}{Emma Renold}.} \bibinfo{year}{1999}\natexlab{}.
\newblock \showarticletitle{The {Use} of {Vignettes} in {Qualitative} {Research}}.
\newblock \bibinfo{journal}{\emph{Social Research Update 25}} \bibinfo{number}{25} (\bibinfo{year}{1999}).
\newblock
\urldef\tempurl%
\url{https://sru.soc.surrey.ac.uk/SRU25.html}
\showURL{%
\tempurl}


\bibitem[Cherrington et~al\mbox{.}(2020)]%
        {cherrington_features_2020}
\bibfield{author}{\bibinfo{person}{Marianne Cherrington}, \bibinfo{person}{David Airehrour}, \bibinfo{person}{Joan Lu}, \bibinfo{person}{Qiang Xu}, \bibinfo{person}{David Cameron-Brown}, {and} \bibinfo{person}{Ihaka Dunn}.} \bibinfo{year}{2020}\natexlab{}.
\newblock \showarticletitle{Features of {Human}-{Centred} {Algorithm} {Design}}. In \bibinfo{booktitle}{\emph{2020 30th {International} {Telecommunication} {Networks} and {Applications} {Conference} ({ITNAC})}}. \bibinfo{pages}{1--6}.
\newblock
\showISBNx{2474-154X}
\urldef\tempurl%
\url{https://doi.org/10.1109/ITNAC50341.2020.9315169}
\showDOI{\tempurl}
\newblock
\shownote{Journal Abbreviation: 2020 30th International Telecommunication Networks and Applications Conference (ITNAC)}.


\bibitem[Coulton and Lindley(2019)]%
        {coulton_more-than_2019}
\bibfield{author}{\bibinfo{person}{Paul Coulton} {and} \bibinfo{person}{Joseph Lindley}.} \bibinfo{year}{2019}\natexlab{}.
\newblock \showarticletitle{More-{Than} {Human} {Centred} {Design}: {Considering} {Other} {Things}}.
\newblock \bibinfo{journal}{\emph{The Design Journal}}  \bibinfo{volume}{22} (\bibinfo{date}{May} \bibinfo{year}{2019}), \bibinfo{pages}{1--19}.
\newblock
\urldef\tempurl%
\url{https://doi.org/10.1080/14606925.2019.1614320}
\showDOI{\tempurl}


\bibitem[Crawford(2021)]%
        {crawford_atlas_2021}
\bibfield{author}{\bibinfo{person}{Kate Crawford}.} \bibinfo{year}{2021}\natexlab{}.
\newblock \bibinfo{booktitle}{\emph{Atlas of {AI}: power, politics, and the planetary costs of artificial intelligence}}.
\newblock \bibinfo{publisher}{Yale University Press}, \bibinfo{address}{New Haven}.
\newblock
\showISBNx{978-0-300-20957-0}
\newblock
\shownote{OCLC: on1111967630}.


\bibitem[{DAIR.AI}(2023)]%
        {dairai_adversarial_2023}
\bibfield{author}{\bibinfo{person}{{DAIR.AI}}.} \bibinfo{year}{2023}\natexlab{}.
\newblock \bibinfo{title}{Adversarial {Prompting}}.
\newblock
\newblock
\urldef\tempurl%
\url{https://www.promptingguide.ai/risks/adversarial}
\showURL{%
\tempurl}


\bibitem[DeVito et~al\mbox{.}(2021)]%
        {devito_queer_2021}
\bibfield{author}{\bibinfo{person}{Michael~Ann DeVito}, \bibinfo{person}{Caitlin Lustig}, \bibinfo{person}{Ellen Simpson}, \bibinfo{person}{Kimberley Allison}, \bibinfo{person}{Tee Chuanromanee}, \bibinfo{person}{Katta Spiel}, \bibinfo{person}{Amy Ko}, \bibinfo{person}{Jennifer Rode}, \bibinfo{person}{Brianna Dym}, \bibinfo{person}{Michael Muller}, \bibinfo{person}{Morgan Klaus~Scheuerman}, \bibinfo{person}{Ashley Marie~Walker}, \bibinfo{person}{Jed Brubaker}, {and} \bibinfo{person}{Alex Ahmed}.} \bibinfo{year}{2021}\natexlab{}.
\newblock \showarticletitle{Queer in {HCI}: {Strengthening} the {Community} of {LGBTQIA}+ {Researchers} and {Research}}. In \bibinfo{booktitle}{\emph{Extended {Abstracts} of the 2021 {CHI} {Conference} on {Human} {Factors} in {Computing} {Systems}}} \emph{(\bibinfo{series}{{CHI} {EA} '21})}. \bibinfo{publisher}{Association for Computing Machinery}, \bibinfo{address}{New York, NY, USA}, \bibinfo{pages}{1--3}.
\newblock
\showISBNx{978-1-4503-8095-9}
\urldef\tempurl%
\url{https://doi.org/10.1145/3411763.3450403}
\showDOI{\tempurl}


\bibitem[Eubanks(2019)]%
        {eubanks_automating_2019}
\bibfield{author}{\bibinfo{person}{Virginia Eubanks}.} \bibinfo{year}{2019}\natexlab{}.
\newblock \bibinfo{booktitle}{\emph{Automating inequality: how high-tech tools profile, police, and punish the poor} (\bibinfo{edition}{first picador edition} ed.)}.
\newblock \bibinfo{publisher}{Picador St. Martin's Press}, \bibinfo{address}{New York}.
\newblock
\showISBNx{978-1-250-21578-9}


\bibitem[Frauenberger and Purgathofer(2019)]%
        {frauenberger_ways_2019}
\bibfield{author}{\bibinfo{person}{Christopher Frauenberger} {and} \bibinfo{person}{Peter Purgathofer}.} \bibinfo{year}{2019}\natexlab{}.
\newblock \showarticletitle{Ways of thinking in informatics}.
\newblock \bibinfo{journal}{\emph{Commun. ACM}} \bibinfo{volume}{62}, \bibinfo{number}{7} (\bibinfo{date}{June} \bibinfo{year}{2019}), \bibinfo{pages}{58--64}.
\newblock
\showISSN{0001-0782}
\urldef\tempurl%
\url{https://doi.org/10.1145/3329674}
\showDOI{\tempurl}


\bibitem[Haxvig(2023)]%
        {haxvig_exploring_2023}
\bibfield{author}{\bibinfo{person}{Helena~A Haxvig}.} \bibinfo{year}{2023}\natexlab{}.
\newblock \showarticletitle{Exploring {Large} {Language} {Model} {Interfaces} {Through} {Critical} and {Participatory} {Design}}. In \bibinfo{booktitle}{\emph{{CHItaly} 2023 {Proceedings} of the {Doctoral} {Consortium} of the 15th {Biannual} {Conference} of the {Italian} {SIGCHI} {Chapter} ({CHItaly} 2023)}}. \bibinfo{address}{Italy}.
\newblock
\urldef\tempurl%
\url{https://ceur-ws.org/Vol-3481/paper4.pdf}
\showURL{%
\tempurl}


\bibitem[Kaltheuner(2021)]%
        {kaltheuner_fake_2021}
\bibfield{author}{\bibinfo{person}{Frederike Kaltheuner}.} \bibinfo{year}{2021}\natexlab{}.
\newblock \bibinfo{booktitle}{\emph{Fake {AI}}}.
\newblock \bibinfo{publisher}{Meatspace Press}.
\newblock
\showISBNx{978-1-913824-02-0}
\newblock
\shownote{OCLC: 1292530708}.


\bibitem[Loi et~al\mbox{.}(2019)]%
        {loi_co-designing_2019}
\bibfield{author}{\bibinfo{person}{Daria Loi}, \bibinfo{person}{Christine~T. Wolf}, \bibinfo{person}{Jeanette~L. Blomberg}, \bibinfo{person}{Raphael Arar}, {and} \bibinfo{person}{Margot Brereton}.} \bibinfo{year}{2019}\natexlab{}.
\newblock \showarticletitle{Co-designing {AI} {Futures}: {Integrating} {AI} {Ethics}, {Social} {Computing}, and {Design}}. In \bibinfo{booktitle}{\emph{Companion {Publication} of the 2019 on {Designing} {Interactive} {Systems} {Conference} 2019 {Companion}}} \emph{(\bibinfo{series}{{DIS} '19 {Companion}})}. \bibinfo{publisher}{Association for Computing Machinery}, \bibinfo{address}{New York, NY, USA}, \bibinfo{pages}{381--384}.
\newblock
\showISBNx{978-1-4503-6270-2}
\urldef\tempurl%
\url{https://doi.org/10.1145/3301019.3320000}
\showDOI{\tempurl}


\bibitem[Mökander et~al\mbox{.}(2023)]%
        {mokander_auditing_2023}
\bibfield{author}{\bibinfo{person}{Jakob Mökander}, \bibinfo{person}{Jonas Schuett}, \bibinfo{person}{Hannah~Rose Kirk}, {and} \bibinfo{person}{Luciano Floridi}.} \bibinfo{year}{2023}\natexlab{}.
\newblock \showarticletitle{Auditing large language models: a three-layered approach}.
\newblock \bibinfo{journal}{\emph{AI and Ethics}} (\bibinfo{date}{May} \bibinfo{year}{2023}).
\newblock
\showISSN{2730-5953, 2730-5961}
\urldef\tempurl%
\url{https://doi.org/10.1007/s43681-023-00289-2}
\showDOI{\tempurl}


\bibitem[Papakyriakopoulos et~al\mbox{.}(2021)]%
        {papakyriakopoulos_qualitative_2021}
\bibfield{author}{\bibinfo{person}{Orestis Papakyriakopoulos}, \bibinfo{person}{Elizabeth~Anne Watkins}, \bibinfo{person}{Amy Winecoff}, \bibinfo{person}{Klaudia Jaźwińska}, {and} \bibinfo{person}{Tithi Chattopadhyay}.} \bibinfo{year}{2021}\natexlab{}.
\newblock \showarticletitle{Qualitative {Analysis} for {Human} {Centered} {AI}}.
\newblock \bibinfo{journal}{\emph{arXiv preprint arXiv:2112.03784}} (\bibinfo{year}{2021}).
\newblock


\bibitem[Rozado(2023)]%
        {rozado_political_2023}
\bibfield{author}{\bibinfo{person}{David Rozado}.} \bibinfo{year}{2023}\natexlab{}.
\newblock \showarticletitle{The {Political} {Biases} of {ChatGPT}}.
\newblock \bibinfo{journal}{\emph{Social Sciences}} \bibinfo{volume}{12}, \bibinfo{number}{3} (\bibinfo{date}{March} \bibinfo{year}{2023}), \bibinfo{pages}{148}.
\newblock
\showISSN{2076-0760}
\urldef\tempurl%
\url{https://doi.org/10.3390/socsci12030148}
\showDOI{\tempurl}
\newblock
\shownote{Number: 3 Publisher: Multidisciplinary Digital Publishing Institute}.


\bibitem[Sampson and Johannessen(2020)]%
        {sampson_turning_2020}
\bibfield{author}{\bibinfo{person}{Helen Sampson} {and} \bibinfo{person}{Idar~Alfred Johannessen}.} \bibinfo{year}{2020}\natexlab{}.
\newblock \showarticletitle{Turning on the tap: the benefits of using ‘real-life’ vignettes in qualitative research interviews}.
\newblock \bibinfo{journal}{\emph{Qualitative Research}} \bibinfo{volume}{20}, \bibinfo{number}{1} (\bibinfo{date}{Feb.} \bibinfo{year}{2020}), \bibinfo{pages}{56--72}.
\newblock
\showISSN{1468-7941}
\urldef\tempurl%
\url{https://doi.org/10.1177/1468794118816618}
\showDOI{\tempurl}
\newblock
\shownote{Publisher: SAGE Publications}.


\bibitem[Scheuerman et~al\mbox{.}(2019)]%
        {scheuerman_how_2019}
\bibfield{author}{\bibinfo{person}{Morgan~Klaus Scheuerman}, \bibinfo{person}{Jacob~M. Paul}, {and} \bibinfo{person}{Jed~R. Brubaker}.} \bibinfo{year}{2019}\natexlab{}.
\newblock \showarticletitle{How {Computers} {See} {Gender}: {An} {Evaluation} of {Gender} {Classification} in {Commercial} {Facial} {Analysis} {Services}}.
\newblock \bibinfo{journal}{\emph{Proceedings of the ACM on Human-Computer Interaction}} \bibinfo{volume}{3}, \bibinfo{number}{CSCW} (\bibinfo{date}{Nov.} \bibinfo{year}{2019}), \bibinfo{pages}{144:1--144:33}.
\newblock
\urldef\tempurl%
\url{https://doi.org/10.1145/3359246}
\showDOI{\tempurl}


\bibitem[Simonsen and Robertson(2013)]%
        {simonsen_routledge_2013}
\bibfield{editor}{\bibinfo{person}{Jesper Simonsen} {and} \bibinfo{person}{Toni Robertson}} (Eds.). \bibinfo{year}{2013}\natexlab{}.
\newblock \bibinfo{booktitle}{\emph{Routledge international handbook of participatory design}}.
\newblock \bibinfo{publisher}{Routledge}, \bibinfo{address}{London}.
\newblock
\showISBNx{978-0-415-69440-7 978-0-415-72021-2 978-0-203-10854-3 978-1-136-26619-5}
\newblock
\shownote{OCLC: 818827037}.


\bibitem[Strengers et~al\mbox{.}(2020)]%
        {strengers_adhering_2020}
\bibfield{author}{\bibinfo{person}{Yolande Strengers}, \bibinfo{person}{Lizhen Qu}, \bibinfo{person}{Qiongkai Xu}, {and} \bibinfo{person}{Jarrod Knibbe}.} \bibinfo{year}{2020}\natexlab{}.
\newblock \showarticletitle{Adhering, {Steering}, and {Queering}: {Treatment} of {Gender} in {Natural} {Language} {Generation}}. In \bibinfo{booktitle}{\emph{Proceedings of the 2020 {CHI} {Conference} on {Human} {Factors} in {Computing} {Systems}}} \emph{(\bibinfo{series}{{CHI} '20})}. \bibinfo{publisher}{Association for Computing Machinery}, \bibinfo{address}{New York, NY, USA}, \bibinfo{pages}{1--14}.
\newblock
\showISBNx{978-1-4503-6708-0}
\urldef\tempurl%
\url{https://doi.org/10.1145/3313831.3376315}
\showDOI{\tempurl}


\bibitem[Wan et~al\mbox{.}(2023)]%
        {wan_kelly_2023}
\bibfield{author}{\bibinfo{person}{Yixin Wan}, \bibinfo{person}{George Pu}, \bibinfo{person}{Jiao Sun}, \bibinfo{person}{Aparna Garimella}, \bibinfo{person}{Kai-Wei Chang}, {and} \bibinfo{person}{Nanyun Peng}.} \bibinfo{year}{2023}\natexlab{}.
\newblock \showarticletitle{“{Kelly} is a {Warm} {Person}, {Joseph} is a {Role} {Model}”: {Gender} {Biases} in {LLM}-{Generated} {Reference} {Letters}}. In \bibinfo{booktitle}{\emph{Findings of the {Association} for {Computational} {Linguistics}: {EMNLP} 2023}}, \bibfield{editor}{\bibinfo{person}{Houda Bouamor}, \bibinfo{person}{Juan Pino}, {and} \bibinfo{person}{Kalika Bali}} (Eds.). \bibinfo{publisher}{Association for Computational Linguistics}, \bibinfo{address}{Singapore}, \bibinfo{pages}{3730--3748}.
\newblock
\urldef\tempurl%
\url{https://doi.org/10.18653/v1/2023.findings-emnlp.243}
\showDOI{\tempurl}


\bibitem[Xu et~al\mbox{.}(2023)]%
        {xu_llm_2023}
\bibfield{author}{\bibinfo{person}{Xilie Xu}, \bibinfo{person}{Keyi Kong}, \bibinfo{person}{Ning Liu}, \bibinfo{person}{Lizhen Cui}, \bibinfo{person}{Di Wang}, \bibinfo{person}{Jingfeng Zhang}, {and} \bibinfo{person}{Mohan Kankanhalli}.} \bibinfo{year}{2023}\natexlab{}.
\newblock \bibinfo{title}{An {LLM} can {Fool} {Itself}: {A} {Prompt}-{Based} {Adversarial} {Attack}}.
\newblock
\newblock
\urldef\tempurl%
\url{https://doi.org/10.48550/arXiv.2310.13345}
\showDOI{\tempurl}
\newblock
\shownote{arXiv:2310.13345 [cs]}.


\end{thebibliography}

\end{document}